\newcounter{myctr}

\documentclass{ws-ijqi}

\begin{document}

\markboth{Paolo Facchi, Giuseppe Florio, Saverio Pascazio} {A
characterization of multipartite entanglement}

\catchline{}{}{}{}{}

\newcommand{\beq}{\begin{equation}}
\newcommand{\eeq}{\end{equation}}
\newcommand{\barr}{\begin{eqnarray}}
\newcommand{\earr}{\end{eqnarray}}

\newcommand{\andy}[1]{ }

\newcommand{\bmsub}[1]{\mbox{\boldmath\scriptsize $#1$}}

\def\bra#1{\langle #1 |}
\def\ket#1{| #1 \rangle}
\def\sinc{\mathop{\text{sinc}}\nolimits}
\def\cV{\mathcal{V}}
\def\cH{\mathcal{H}}
\def\cT{\mathcal{T}}
\renewcommand{\Re}{\mathop{\text{Re}}\nolimits}
\newcommand{\tr}{\mathop{\text{Tr}}\nolimits}

\title{CHARACTERIZING AND MEASURING MULTIPARTITE ENTANGLEMENT}

\author{P. FACCHI}

\address{Dipartimento di Matematica, Universit\`a di Bari, I-70125  Bari, Italy\\
Istituto Nazionale di Fisica Nucleare, Sezione di Bari, I-70126
Bari, Italy}

\author{G. FLORIO and S. PASCAZIO}

\address{Dipartimento di Fisica, Universit\`a di Bari,
I-70126  Bari, Italy\\
Istituto Nazionale di Fisica Nucleare, Sezione di Bari, I-70126
Bari, Italy}

\maketitle


\begin{abstract}
A method is proposed to characterize and quantify multipartite
entanglement in terms of the probability density function of
bipartite entanglement over all possible balanced bipartitions of an
ensemble of qubits. The method is tested on a class of random pure
states.
\end{abstract}

\vspace{.4cm}

The quantification of multipartite entanglement is an open and very
challenging problem. An exhaustive definition of
\emph{bipartite} entanglement exists and hinges upon the von Neumann entropy
and the entanglement of formation,\cite{woot} but the problem of
defining \emph{multipartite} entanglement is more
difficult\cite{druss} and no unique definition exists: different
definitions tend indeed to focus on different aspects of the
problem, capturing different features of
entanglement,\cite{multipart} that do not necessarily agree with
each other. Moreover, as the size of the system increases, the
number of measures (i.e.\ real numbers) needed to quantify
multipartite entanglement grows exponentially.

This work is motivated by the idea that a good definition of
multipartite entanglement should stem from some statistical
information about the system.\cite{FFP} We shall therefore look at
the distribution of the purity of a subsystem over all bipartitions
of the total system. As a measure of multipartite entanglement we
will take a whole \emph{function}: the probability density of
bipartite entanglement between any two parts of the total system.
According to our definition multipartite entanglement is large when
bipartite entanglement (i) is large \emph{and} (ii) does not depend
on the bipartition, namely when (i$+$ii) the probability density of
bipartite entanglement is a narrow function centered at a large
value. This definition will be tested on two class of states that
are known to be characterized by a large entanglement. We emphasize
that the idea that complicated phenomena cannot be ``summarized" in
a single (or a few) number(s) was already proposed in the context of
complex systems\cite{parisi} and has been also considered in
relation to quantum entanglement.\cite{MMSZ}

We shall focus on a collection of $n$ qubits and consider a
partition in two subsystems $A$ and $B$, made up of $n_A$ and $n_B$
qubits ($n_A+n_B=n$), respectively. For definiteness we assume $n_A
\le n_B$. The total Hilbert space is the tensor product
$\mathcal{H}=\mathcal{H}_A\otimes\mathcal{H}_B$ and the dimensions
are $\dim \mathcal{H}=N=2^n$, $\dim \mathcal{H}_A=N_A=2^{n_A}$ and
$\dim \mathcal{H}_B=N_B=2^{n_B}$, respectively ($N_AN_B=N$).

We shall consider pure states. Their expression adapted to the
bipartition reads
\begin{equation}
|\psi\rangle = \sum_{k=0}^{N-1} z_k |k\rangle = \sum_{j_A=0}^{N_A-1}
\sum_{l_B=0}^{N_B-1} z_{j_A l_B} |j_A\rangle\otimes|l_B\rangle ,
\label{eq:genrandomx}
\end{equation}
where $|k\rangle=|j_A\rangle\otimes|l_B\rangle$, with a bijection
between $k$ and $(j_A,l_B)$. Think of the binary expressions of an
integer $k$ in terms of the binary expression of $(j_A, l_B)$.

As a measure of \emph{bipartite} entanglement between $A$ and $B$ we
consider the participation number
\begin{equation}\label{eq:NAB}
N_{AB}=\pi_{AB}^{-1},\qquad \pi_{AB}(|\psi\rangle)=\mathrm{tr}_A
\rho_A^2, \qquad \rho_A=\mathrm{tr}_B \rho,
\end{equation}
where $\rho=|\psi\rangle\langle\psi|$ and $\mathrm{tr}_A$
($\mathrm{tr}_B$) is the partial trace over the subsystem $A$ ($B$).
$N_{AB}$ measures the effective rank of the matrix $\rho_A$, namely
the effective Schmidt number.\cite{eberly} We note that
\begin{equation}\label{eq:propNAB}
1\leq N_{AB}=N_{BA}\leq \text{min}(N_A,N_B),
\end{equation}
with the maximum (minimum) value attained for a completely mixed
(pure) state $\rho_A$. Therefore, a larger value of $N_{AB}$
corresponds to a more entangled bipartition $(A,B)$, the maximum
value being attainable for a \emph{balanced} bipartition, i.e.\ when
$n_A = [n/2]$ (and $n_B=[(n+1)/2]$), where $[x]$ is the integer part
of the real $x$, that is the largest integer not exceeding $x$. The
maximum possible entanglement is $N_{AB} = N_A= 2^{n_A}$. The
quantity $n_{AB}= \log_2 N_{AB}$ represents the effective number of
entangled qubits, given the bipartition (namely, the number of
bipartite entanglement ``links" that are ``severed" when the system
is bipartitioned).

Clearly, the quantity $N_{AB}$ will depend on the bipartition, as in
general entanglement will be distributed in a different way among
all possible bipartitions. As explained in the introduction, we are
motivated by the idea that the distribution $p(N_{AB})$ of $N_{AB}$
yields information about \emph{multipartite} entanglement.

Let us therefore study the typical form of our measure of
multipartite entanglement $p(N_{AB})$ for a very large class of pure
states, sampled according to a given symmetric distribution on the
projective Hilbert space $\{\psi\in\cH ,
\|\psi\|=1\}$ (e.g. the unitarily invariant Haar measure).
By plugging (\ref{eq:genrandomx}) into (\ref{eq:NAB}) one gets
\begin{equation}\label{eq:piAB}
\pi_{AB}=\sum_{j,j'=0}^{N_A-1}\,\sum_{l,l'=0}^{N_B-1}z_{j l} \bar
z_{j' l} z_{j' l'} \bar z_{j l'}
\end{equation}
and it can be shown\cite{FFP} that, in the thermodynamical limit
(that is practically attained for $n > 5$), independently of the
distribution of the coefficients, the mean and the standard
deviation of (\ref{eq:piAB}) over all possible balanced bipartitions
read
\beq \mu_{AB}
= \frac{N_A+N_B-1}{N} = \sqrt{\frac{\alpha}{N}} \, ,
\quad
\sigma^2_{AB}=\frac{2}{N^2},
\quad (N \; \mbox{large})
\label{eq:muABas}
\eeq
respectively, where $\alpha=8/2$ $(\alpha=9/2)$ for even (odd) $n$.
Moreover, the probability density of $N_{AB}$ in
Eq.~({\ref{eq:NAB}}) reads
\begin{eqnarray}
p(N_{AB})&=&\frac{1}{N^2_{AB}(2\pi\sigma_{AB}^2)^{1/2}}
\exp\left(-\frac{(N_{AB}^{-1}-\mu_{AB})^2}{2\sigma_{AB}^2}\right) .
\label{eq:distrpartnumber}
\end{eqnarray}
It is interesting to compare the features of these generic random
states with those of other states studied in the literature.
\begin{table}[ph]
\tbl{\label{tab:confront}Mean bipartite entanglement $\langle
N_{AB}\rangle$ for different states and different number of qubits
$n$.} {\begin{tabular}{cc|ccccc}
\toprule & $n$ & GHZ
& W & cluster & random &\\
 \colrule
& 5 & 2 & 1.923 & 3.6 & 2.909 &  \\

&6 & 2 & 2 & 5.4 & 4.267 \\

&7 & 2 & 1.96 & 6.171 & 5.565 \\

&8 & 2 & 2 & 8.743 & 8.258 \\

&9 & 2 & 1.976 & 10.349 & 10.894 \\

&10 & 2 & 2 & 14.206 & 16.254 \\

&11 & 2 & 1.984 & 17.176 & 21.558 \\

& 12 & 2 & 2 & 23.156 & 32.252\\ \botrule
\end{tabular}}
\end{table}
Table \ref{tab:confront} displays the average value of $N_{AB}$ for
GHZ states,\cite{ghz} W states,\cite{w} cluster states\cite{briegel}
and the generic states (\ref{eq:genrandomx}), for $n=5\div12$. While
the entanglement of the GHZ and W states is essentially independent
of $n$, the situation is drastically different for cluster and
random states. In both cases, the average entanglement increases
with $n$; for $n>8$ the average entanglement is higher for random
states. However, the mean $\langle N_{AB}\rangle$ yields poor
information on multipartite entanglement. For this reason, it is
useful to analyze the distribution of bipartite entanglement over
all possible balanced bipartitions.
\begin{figure}
\includegraphics[width=0.99 \textwidth]{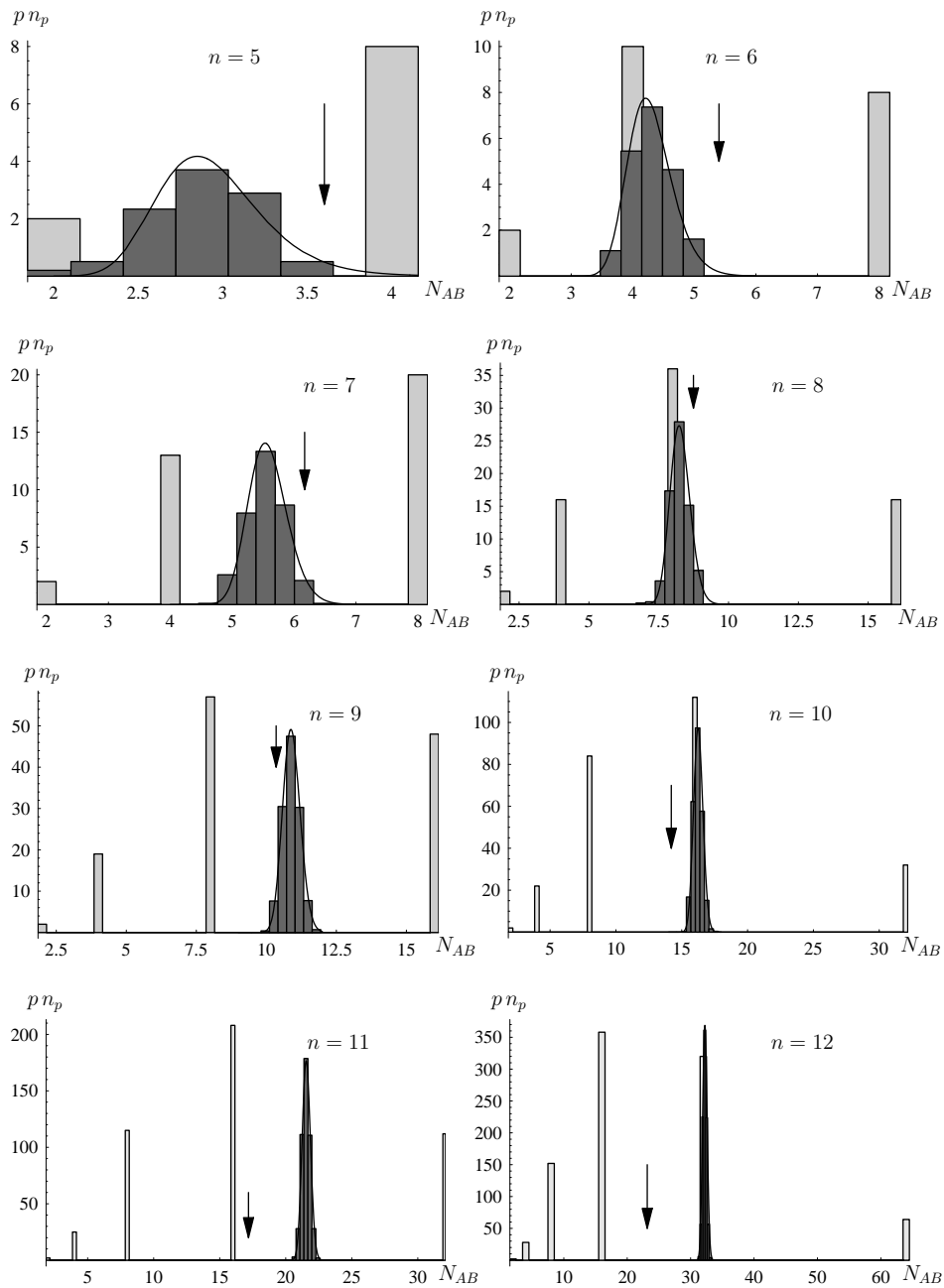}
\caption{Number of balanced bipartitions vs $N_{AB}$;
$n_p=n!/n_A!n_B!$ is the number of bipartitions. the light-gray bars
represent cluster states, the dark-gray ones random states; the
solid line is the distribution
(\ref{eq:muABas})-(\ref{eq:distrpartnumber}); the black arrows
indicate the average $\langle N_{AB}\rangle_{\text{cluster}}$. For
even $n$ ($n=12$ in particular) the distribution of the random state
partially hides a peak of the corresponding cluster state
distribution, centered at $N_{AB}= 2^{n_A-1} = 2^{[n/2]-1}$.}
\label{confrontorandomclusternew}
\end{figure}

The results for the cluster and random states are shown in Fig.\
\ref{confrontorandomclusternew}, for $n$ ranging between 5 and 12. Notice
that the distribution function of the random state is always peaked
around $\langle N_{AB} \rangle \simeq \mu_{AB}^{-1}$ given by
(\ref{eq:muABas}). Notice also that the cluster state can reach
higher values of $N_{AB}$ (the maximum possible value being
$2^{[n/2]}$), however, the fraction of bipartitions yielding this
result becomes smaller for higher $n$. This is immediately
understood if one realizes that the cluster states are designed for
optimized applications and therefore perform better in terms of
\emph{specific} bipartitions. On the other hand, according to the measure
we propose, the random states are characterized by a large value of
multipartite entanglement, that is roughly \emph{independent} of the
bipartition.

\begin{figure}
\begin{center}
\includegraphics[width=1 \textwidth]{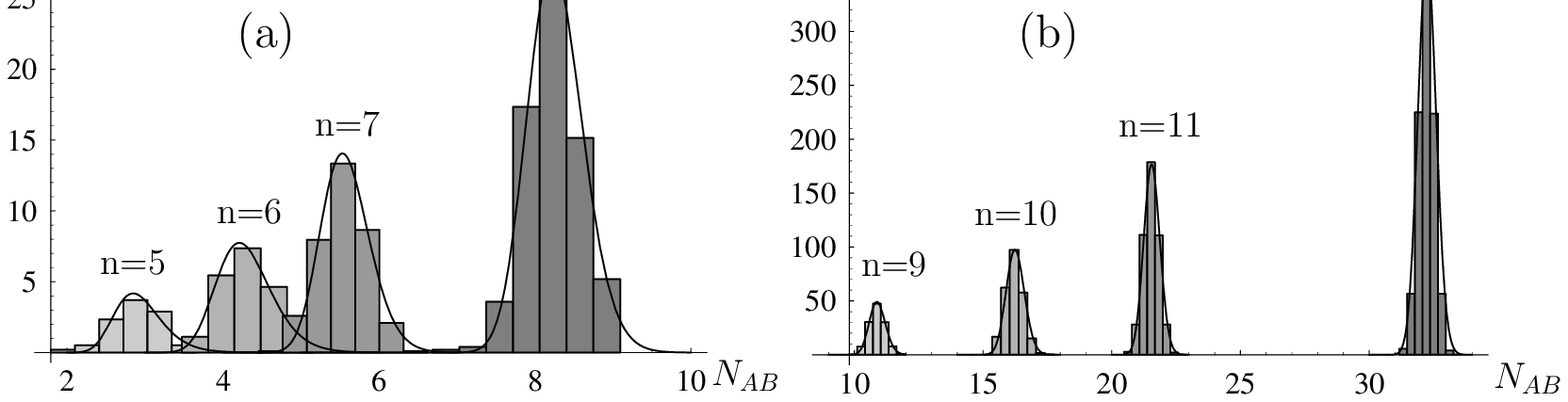}
\caption{Number of balanced bipartitions vs $N_{AB}$; the
histograms are numerically obtained for the typical states; the
solid line represents their distribution. The value of $n$ (total
number of spins) is always indicated and ranges (a) from 5 to 8; (b)
from 9 to 12; $n_p=n!/n_A!n_B!$ is the number of bipartitions. }
\label{pic512}
\end{center}
\end{figure}

\begin{figure}
\begin{center}
\includegraphics[width=0.7 \textwidth]{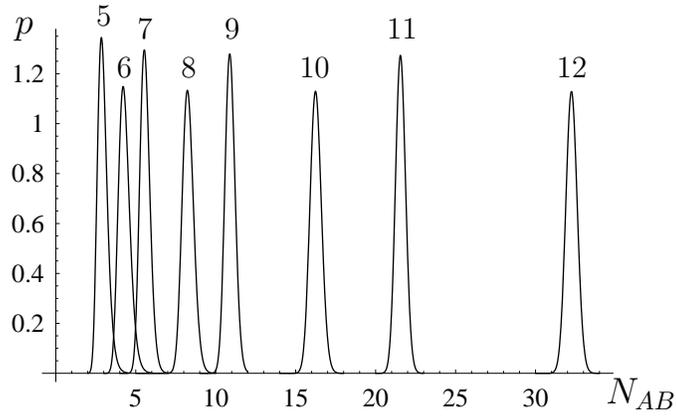}
\caption{Probability densities functions
(\ref{eq:distrpartnumber}) vs $N_{AB}$. Each curve is labeled with
the corresponding value of $n$ (number of qubits). The standard
deviation $\sigma$ quickly becomes independent of $n$ [see Fig.\
\ref{sigmansite}(b)] and depends only on parity of the latter.}
\label{picprob}
\end{center}
\end{figure}

In Fig.\ \ref{pic512} we compare the number of balanced bipartitions
vs $N_{AB}$ for the random states and increasing $n$. The related
probability density functions (\ref{eq:distrpartnumber}) are
displayed in Fig.\ \ref{picprob}. Notice that as the number of spins
increases from $n=5$ to $n=12$ the mean increases and the
distribution becomes relatively narrower. As we emphasized, these
are both signatures of a very high degree of multipartite
entanglement, whose features become (as $n$ increases) practically
independent of the bipartition. In Fig.\ \ref{picprob} it is
interesting to observe the difference between the distributions for
odd and even $n$.

\begin{figure}
\begin{center}
\includegraphics[width=1 \textwidth]{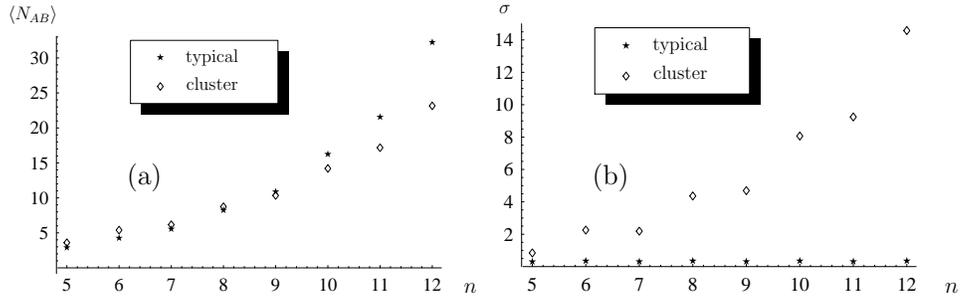}
\caption{Comparison between typical and cluster states:
(a) expectation value $\langle N_{AB}\rangle$ and (b) standard
deviation $\sigma$ of the distributions in Fig.\ \ref{picprob} vs
$n$ (number of qubits).}
\label{sigmansite}
\end{center}
\end{figure}

\begin{figure}
\begin{center}
\includegraphics[width=0.7 \textwidth]{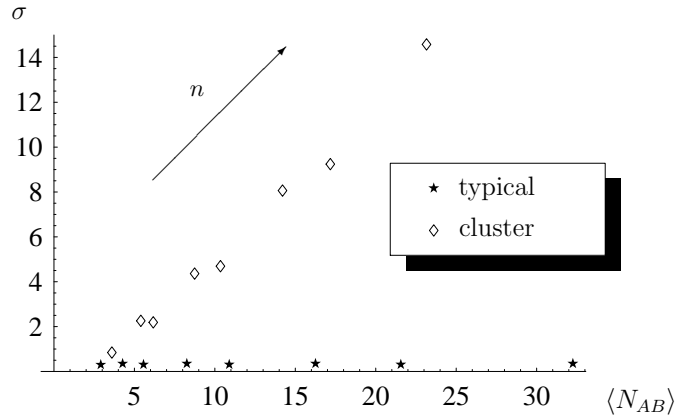}
\caption{Standard deviation $\sigma$ vs expectation value $\langle N_{AB}
\rangle$ of the distributions in Fig.\ \ref{picprob}.} \label{sigman}
\end{center}
\end{figure}

In Fig.\ \ref{sigmansite}(a) we plot the value of $\langle
N_{AB}\rangle$ for the cluster and random states (see Table
\ref{tab:confront}). We notice that, for $n=9$, $\langle
N_{AB}\rangle_{\text{random}}$ becomes larger than $\langle
N_{AB}\rangle_{\text{cluster}}$. Figure
\ref{sigmansite}(b) displays the behavior of the standard deviation
of $N_{AB}$,
\begin{equation}\label{eq:sigma}
\sigma\simeq \sigma_{AB}/\mu_{AB}^2 .
\end{equation}
For the cluster states this quantity tends to diverge when the size
of the system increases. By contrast, from Eq.\ (\ref{eq:muABas}),
$\sigma=\sqrt{2}/\alpha$ is constant for the typical states. This
means that the ratio $\sigma/\langle N_{AB}\rangle$ tends to 0.
Finally, Figure \ref{sigman} displays a parametric plot of $\sigma$
vs $\langle N_{AB}\rangle$. Clearly, for the random states $\sigma$
is independent of $\langle N_{AB}
\rangle$.

We emphasize that our analysis should by no means be taken as an
argument against the performance of the cluster states. As we
stressed before, cluster states are tailored for specific purposes
in quantum information processing, and in that respect are very well
suited. We compared the generic states to the cluster states
specifically because the latter are also known to be characterized
by a large entanglement.

An efficient way to generate states endowed with random features is
by means of a chaotic dynamics,\cite{entvschaos} or at the onset of
a quantum phase transition.\cite{QPT} In particular, the random
states describe quite well states with support on chaotic regions of
phase space, before dynamical localization has taken place. These
features make these states rather appealing, from a practical point
of view, in that they are easily generated. The introduction of a
probability density function as a measure of multipartite
entanglement paves the way to further investigations of the intimate
relation between entanglement and randomness and their behavior
across a phase transition.

\section*{Acknowledgments}

This work is partly supported by the bilateral Italian--Japanese
Projects II04C1AF4E on ``Quantum Information, Computation and
Communication'' of the Italian Ministry of Instruction, University
and Research and by the European Community through the Integrated
Project EuroSQIP.


\end{document}